\documentstyle[12pt,epsf]{article}
\input epsf.sty
\begin{document}
\thispagestyle{empty}
\baselineskip 18pt
%%\pubnum{September 1994 ITP UWr 880/94}
%\input{catmac}
\title{ Properties of an equilibrium hadron gas subjected to the adiabatic
longitudinal expansion
\thanks{ Work partially supported by the Polish Committee for Scientific
Research
under contract KBN-200579101} }
\author{\em Dariusz Prorok and Ludwik Turko
\\ Institute of Theoretical Physics, University
of Wroc{\l}aw,\\ Pl.Maksa Borna 9, 50-204  Wroc{\l}aw,
Poland}
\date{}
\maketitle
\newcommand{\RRR}{R\hspace{-0.4cm}R}
\newcommand{\NNN}{N\hspace{-0.31cm}N}
\newcommand{\ZZZ}{Z\hspace{-0.4cm}Z}
\newcommand{\ds}{\displaystyle}
\newcommand{\un}{\underline}
\newcommand{\be}{\begin{equation}}
\newcommand{\ee}{\end{equation}}
\newcommand{\ba}{\begin{array}}
\newcommand{\ea}{\end{array}}
\newcommand{\bd}{\begin{description}}
\newcommand{\ed}{\end{description}}
\newcommand{\po}{Poincar\'{e}}
\def\NN{{\cal N}}
%\vspace{5cm}
%\newpage
\begin{abstract}
We consider an ideal gas of massive hadrons in thermal and chemical
equilibrium. The gas expands longitudinally in an adiabatic way. This
evolution for a baryonless gas reduces to a hydrodynamic expansion.
Cooling process is parametrized by the sound velocity. The sound
velocity is temperature dependent and is strongly influenced by
hadron mass spectrum.
\end{abstract}
%\vspace{10cm}
%\hrule
%\vspace{0.5cm}

\newpage

\section { Introduction }
In our last paper [1], there have been presented results for $J/\Psi$
suppression in an equilibrium hadron gas. The gas expands longitudinally
according to the Bjorken pattern [2] and its initial conditions have been
estimated from the data [3,4]. Since our purpose in Ref.1. was to study
$J/\Psi$ suppression patterns, we left aside more detailed description
of properties of the hadron gas subjected to the longitudinal cooling.
Now, we would like to discuss this subject more carefully.

In the case of hadron-hadron collisions we shall assume an existence of
a "central region"  in the rapidity variable [2]. We are leaving aside
here a problem of an appearance of a quark-gluon plasma phase. We begin
our analysis with the formation of an equilibrium hadron gas, subjected
to the longitudinal expansion. This gas consists of different species
of hadrons and it has a non-vanishing baryon-number. This fact
corresponds to present experimental conditions. A  neutral central
region is expected for higher then CERN collision energies of heavy
ions.
\section{Equation of state}
An equation of state is given by an assumption of a thermal and
chemical equilibrium of an ideal hadronic gas. Particle ratios are
given by the temperature and chemical potentials related to conserved
quantum numbers -- strangeness and baryon-number. In our model we take
into account all species of hadrons up to $\Omega^{-}$ baryon.

For an ideal hadron gas in thermal and chemical equilibrium, which consists of
$l$ species of particles, energy density $\epsilon$, baryon-number density
 $n_{B}$, strangeness density $n_{S}$ and entropy density $s$ read
($\hbar=c=1$ always)
$$
\epsilon = { 1 \over {2\pi^{2}}} \sum_{i=1}^{l} (2s_{i}+1) \int_{0}^{\infty}
{ { dpp^{2}E_{i} }
\over { \exp \left\{ {{ E_{i} - \mu_{i} } \over T} \right\}
+ g_{i} } } \ ,
\eqno(1a)
$$
$$
n_{B}={ 1 \over {2\pi^{2}}} \sum_{i=1}^{l} (2s_{i}+1) \int_{0}^{\infty}
{ { dpp^{2}B_{i} }
\over { \exp \left\{ {{ E_{i} - \mu_{i} } \over T} \right\}
+ g_{i} } } \ ,
\eqno(1b)
$$
$$
n_{S}={1 \over {2\pi^{2}}} \sum_{i=1}^{l} (2s_{i}+1) \int_{0}^{\infty}
{ { dpp^{2}S_{i} }
\over { \exp \left\{ {{ E_{i} - \mu_{i} } \over T} \right\}
+ g_{i} } } \ ,
\eqno(1c)
$$
$$
s={1 \over {6\pi^{2}T^{2}} } \sum_{i=1}^{l} (2s_{i}+1) \int_{0}^{\infty}
{ {dpp^{4}} \over { E_{i} } } { { (E_{i}
- \mu_{i}) \exp \left\{ {{ E_{i} - \mu_{i} } \over T}
\right\} } \over { \left( \exp \left\{ {{ E_{i} - \mu_{i} }
\over T} \right\} + g_{i} \right)^{2} } }\ ,
\eqno(1d)
$$
where $E_{i}= ( m_{i}^{2} + p^{2} )^{1/2}$ and
$m_{i}$, $B_{i}$, $S_{i}$, $\mu_{i}$, $s_{i}$ and $g_{i}$ are the mass,
baryon-number, strangeness, chemical potential, spin and
a statistical factor of specie $i$ respectively (we treat
an antiparticle as a different specie).

And $\mu_{i} = B_{i}\mu_{B} + S_{i}\mu_{S}$, where $\mu_{B}$ and
$\mu_{S}$ are overall baryon-number and strange\-ness chemical potentials
res\-pe\-cti\-vely.

These equations can be supplemented by a similar equation related to
the electric charge density. A chemical potential $\mu_{Q}$  would be
introduced due to the electric charge conservation. However, it is easy
to see that for the ratio $A/Z = 2$ this chemical potential is always
equal to zero. So we can omit an equation for the electric charge
density at present experiments.
\section{Time evolution}
In general, all densities of the left sides of eqs.1a-d are functions
of time during the cooling of the hadron gas. This means that gas
parameters such as temperature and chemical potentials should be
functions of time. We would like to obtain these functions explicite.
This can be done by solving numerically the system of equations for
entropy density $s$, baryon-number density $n_{B}$ and strangeness
density $n_{S}$, with $s$, $n_{B}$ and $n_{S}$ given as time dependent
quantities. Strangeness density is equal to zero during all the evolution
process, but for other quantities some dynamical assumption is needed.
In the following we shall assume a longitudinal expansion and we shall
neglect transversal degrees of freedom.

{}From the baryon-number conservation we have then
$$
n_{B}(t)= { {n_{B}^{0}t_{0}} \over t } \ ,
\eqno(2a)
$$
{}From the Bjorken model with vanishing baryon-number density we have the
following solution for the longitudinal expansion [1,5]
$$
s(t)= { {s_{0}t_{0}} \over t } \,
\eqno(2b)
$$
where $s_{0}$ and $n_{B}^{0}$ are initial densities of the entropy and the
baryon-number respectively.
This gives an entropy conservation during the expansion stage. We shall
assume an adiabatic cooling process, so eq.2a will be taken for granted
also with nonvanishing baryon-number density. This means a deviation
from a simple hydrodynamical model.

To solve (1b,c,d) with $s$ and $n_{B}$ given by (2) and $n_{S}=0$, we
need to know initial values $s_{0}$ and $n_{B}^{0}$.  To estimate
initial baryon-number density $n_{B}^{0}$ we use experimental results
of [3]. These results are for S-S collisions, but since there are no
data on baryon multiplicities for heavier nuclei we have to evaluate
them in some way.

We assume that the baryon multiplicity per unit rapidity in the CRR is
proportional to the number of participating nucleons. For a
sulphur-sulphur collision we have $dN_{B}/dy \cong 6$ [3] and 64
participating nucleons.  For an O-U collision we can roughly estimate
the number of participating nucleons at $16+58=74$.

The second factor of the sum has been obtained by the following
assumption: since an oxygen nucleon is much smaller than an uranium
one, we can approximate the part of the uranium, through which the
oxygen passes, by the cylinder of the volume equal to $\pi R_{O}^{2}
\cdot 2R_{U}$.  The same procedure can be applied to the S-U case. Here
we obtain $32+93=125$ participants.

Therefore, we have $dN_{B}/dy \cong 7$ and $dN_{B}/dy \cong 11.7$ for
O-U and S-U collisions respectively.  Having taken the initial volume
in the CRR equal to $\pi R_{A}^{2} \cdot 1$ fm, we arrive at $n_{B}^{0}
\cong 0.25 \; {\rm fm^{-3}}$ for both cases.

To find $s_{0}$, first we have to solve (1a,b,c) with respect to $T$,
$\mu_{S}$ and $\mu_{B}$, where we put $\epsilon =
\epsilon_{0}$ and so on.  For $\epsilon_{0}$ we have taken estimates
given in [4].  As a result, we have obtained $T_{0} \cong 212$ MeV,
$\mu_{S}^{0} \cong 34.6$ MeV and $\mu_{B}^{0} \cong 133$ MeV for the
O-U collision ($\epsilon_{0}= 2.5 \; {\rm GeV/fm^{3}}$) and $T_{0}
\cong 209.1$ MeV, $\mu_{S}^{0} \cong 36.7$ MeV and $\mu_{B}^{0} \cong
143.3$ MeV for the S-U collision ($\epsilon_{0}= 2.3 \; {\rm
GeV/fm^{3}}$). Then, from (1d) we have $s_{0} \cong 13.68 \; {\rm
fm^{-3}}$ for O-U and $s_{0} \cong 12.74 \; {\rm fm^{-3}}$ for S-U.

Now, having put (2) and $n_{S}= 0$ into (1b,c,d), we can solve them
numerically to obtain $T$, $\mu_{S}$ and $\mu_{B}$ as functions of
time.
\section{Results}
Our results are presented in fig.1, 2, where solid, long-dashed and
dashed lines mean the temperature, the strangeness chemical potential
and the baryon chemical potential respectively. Fig.1 shows results for
O-U collision initial conditions and fig.2 for S-U ones.  The time
scale is chosen in a way which enables the temperature to reach the
freez-out at 140 MeV (figs.1a and 2) or 100 MeV (fig.1b).  This
corresponds to the freez-out time equal to $t_{f.o.} \cong 10.4$ fm or
$t_{f.o.} \cong 58.6$ fm respectively. The most interesting feature of
our results is the behaviour of the temperature.  For figs.1a, 1b, and
2 the following temperature approximations hold respectively: $$ T(t)
\cong 212.4 \cdot
\left( {1 \over t} \right)^{0.178}
\eqno(3a)
$$
$$
T(t) \cong 217.2 \cdot \left( {1 \over t} \right)^{0.189}
\eqno(3b)
$$
$$
T(t) \cong 209.7 \cdot \left( {1 \over t} \right)^{0.179}.
\eqno(3c)
$$
We can see that all above expressions have the form known from the solution
for the longitudinal expansion of a baryonless gas with the sound velocity
constant, namely [5]
$$
T(t) = T_{0} \cdot \left( {1 \over t} \right)^{ c_{s}^{2} }\ ,
\eqno(4)
$$
where $c_{s}$ is the sound velocity and we put the initial time $t_{0}$
equal to 1 fm.

We have checked that for $n_{B} = 0$ results for the temperature function
are very similar to those in eq.3. This is shown in figs.3 and 4. Fig.3
has the same initial conditions as fig.1b and fig.4 the same as fig.2.
The following approximations of the temperature function hold
$$
T(t) = 215.2 \cdot \left( {1 \over t} \right)^{0.180}
\eqno(5a)
$$
$$
T(t) = 208.9 \cdot \left( {1 \over t} \right)^{0.172}.
\eqno(5b)
$$
We can see that the power in eqs.3 and 4 is around 0.18. Therefore a
question arises: is this value connected in any way with the sound velocity
of the hadron gas? For the baryonless case the answer is positive. We have
checked this computing straightforward

$$c_{s}^{2}=s/(T { {\partial s} \over{\partial T} }) \,$$

which is the value of the sound velocity squared for a baryonless gas
[5]. The results are presented in fig.5.
For comparison, the square of the sound velocity of a
pion gas is also depicted (dashed lines). We can see that in the range
of the temperature 200-140 MeV the square of the speed of sound equals
0.17-0.18 indeed. We have also found a region of the temperature where
the sound velocity decreases as the temperature increases.

Nevertheless, the condition for the stability of the expansion [6],
$d/dT(sc_{s}/T) > 0$ is still valid because $sc_{s}/T$ is an increasing
function of temperature in the whole temperature region.
This function is presented in fig.6.

\newpage

\section*{Figure Captions}

\bd

\item{Fig.1a.} Dependence of the temperature (solid line), the strangeness
chemical potential (long-dashed line) and the baryon chemical potential
(dashed line) on time. Initial conditions are chosen for O-U collision with
freez-out at 140MeV.

\item{Fig.1b.} Same as fig.1a but for freez-out at 100 MeV.

\item{Fig.2.} Same as fig.1a but for S-U collision.

\item{Fig.3.} Dependence of the temperature on time for hadron gas with
non-vanishing (solid line) and vanishing (dashed line) baryon number.
Initial conditions are chosen for O-U collision with freez-out at 100MeV.

\item{Fig.4.} Dependence of the temperature on time for hadron gas with
non-vanishing (solid line) and vanishing (dashed line) baryon number.
Initial conditions are chosen for S-U collision with freez-out at 140MeV.

\item{Fig.5.} Baryonless gas: dependence of the sound velocity squared on
time. Solid line: all hadrons up to $\Omega^{-}$ resonance. Dashed line: pion
gas.

\item{Fig.6.} Stability factor $sc_{s}/T$ for a baryonless hadronic gas.

\ed
\newpage
{\epsfbox{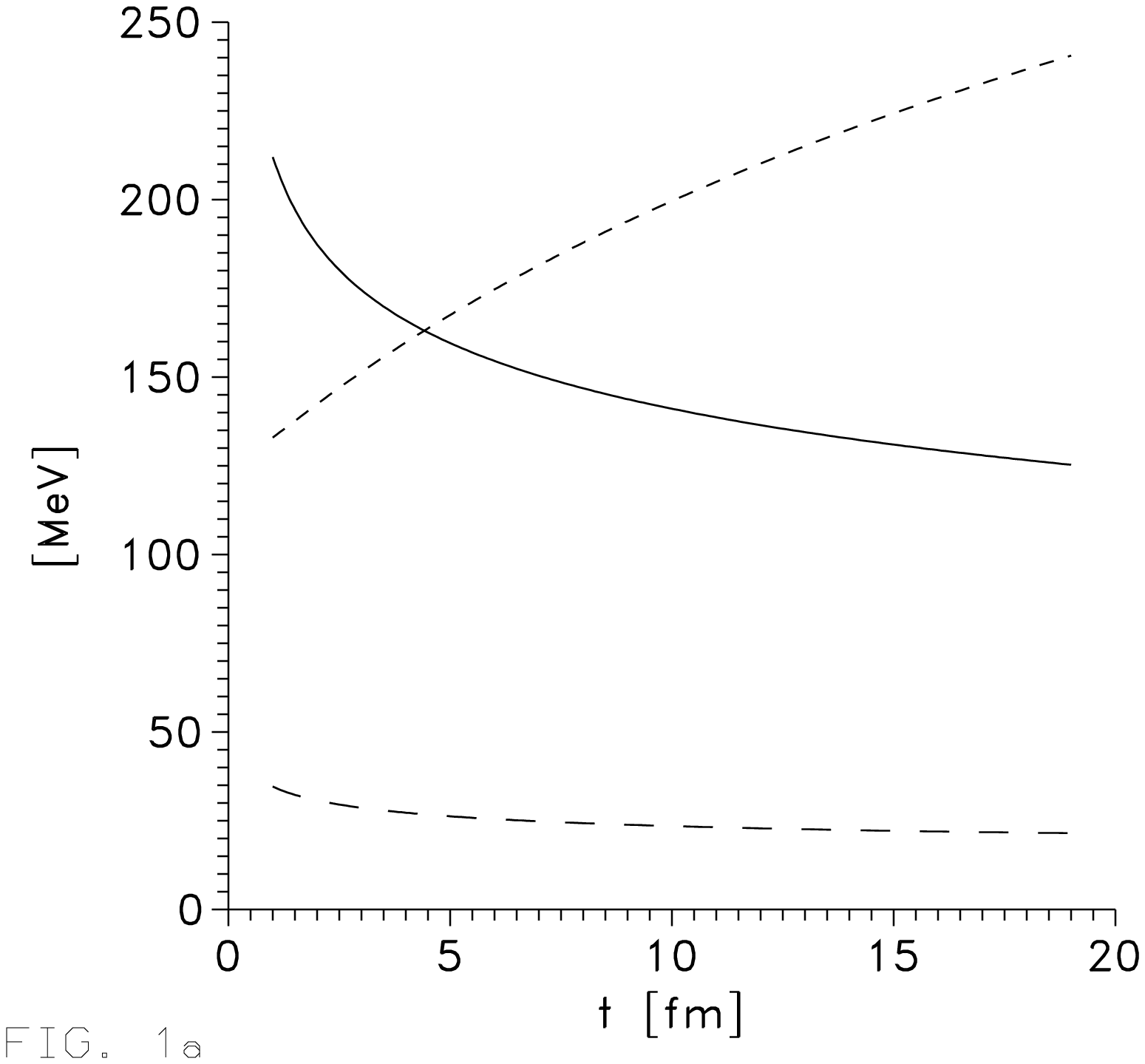}}

\newpage
{\epsfbox{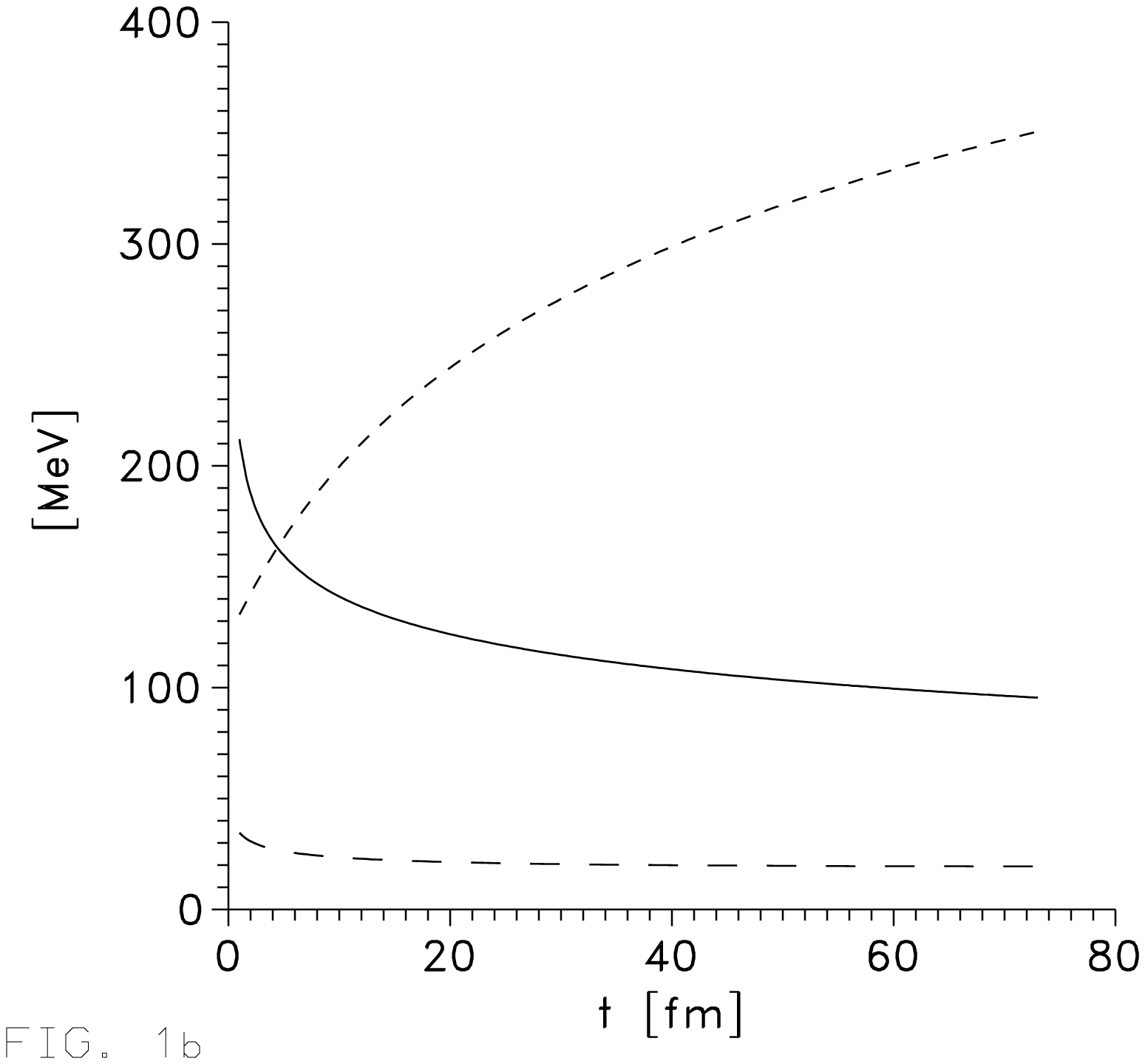}}

\newpage
{\epsfbox{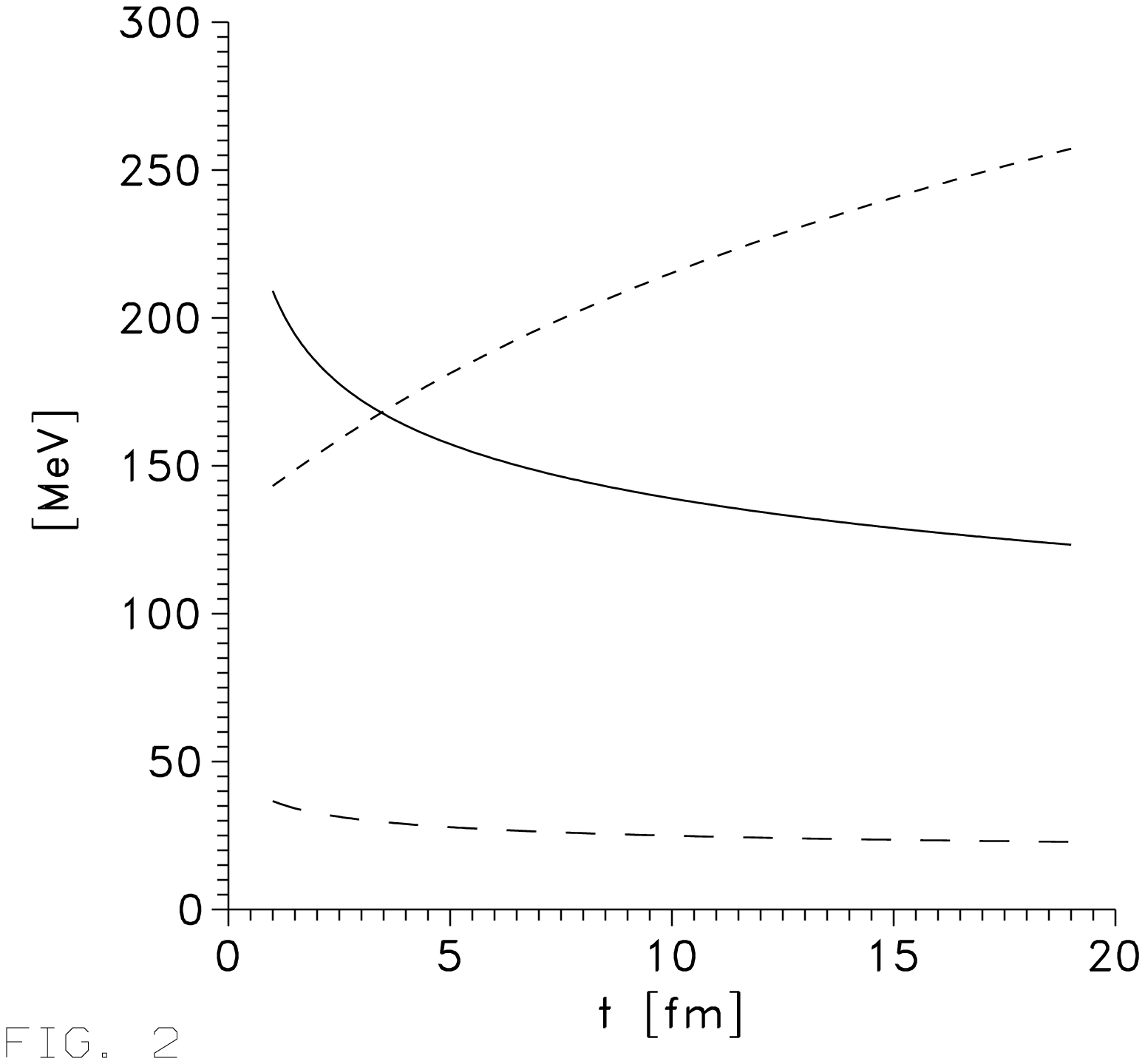}}

\newpage
{\epsfbox{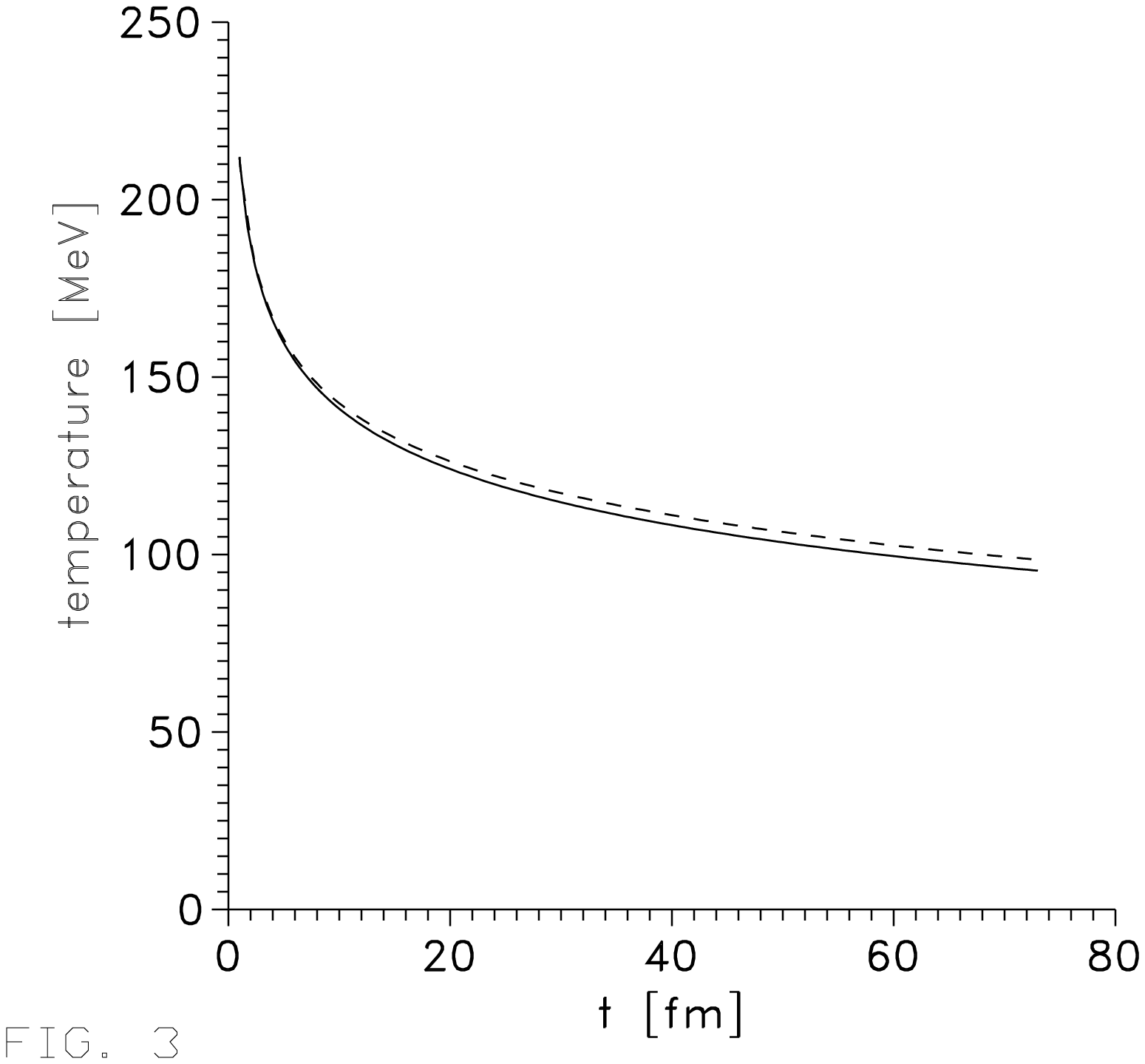}}

\newpage
{\epsfbox{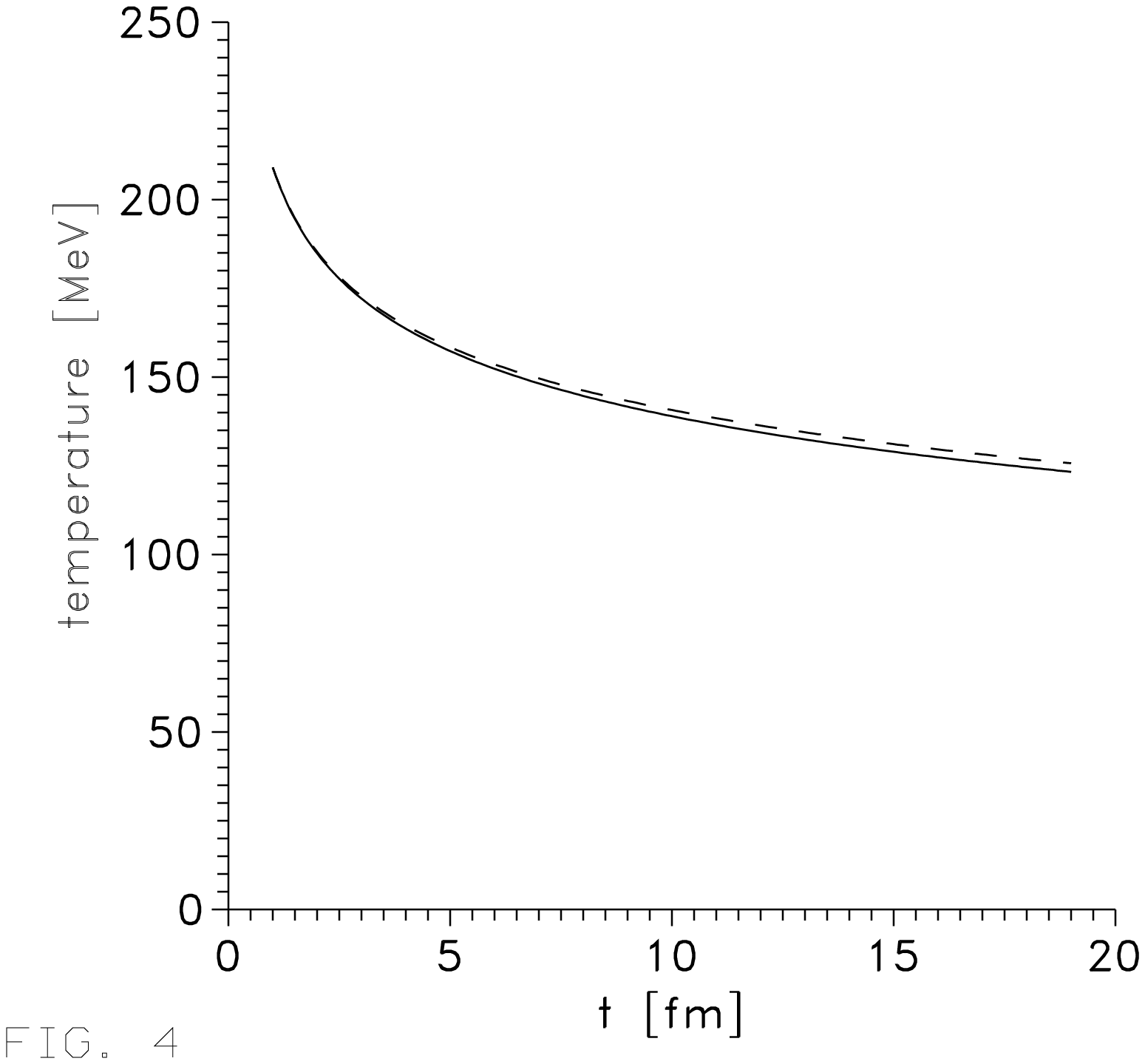}}

\newpage
{\epsfbox{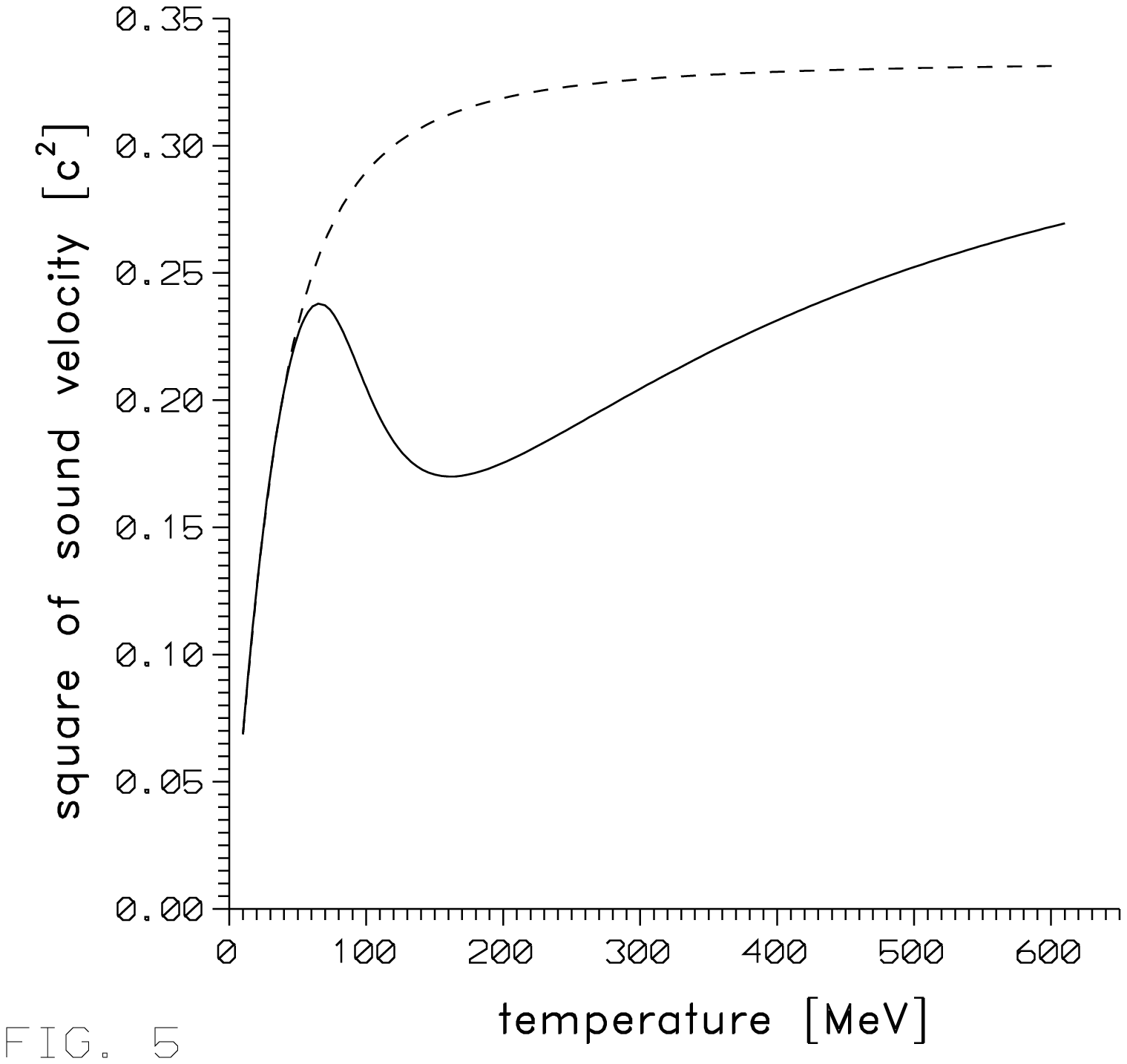}}

\newpage
{\epsfbox{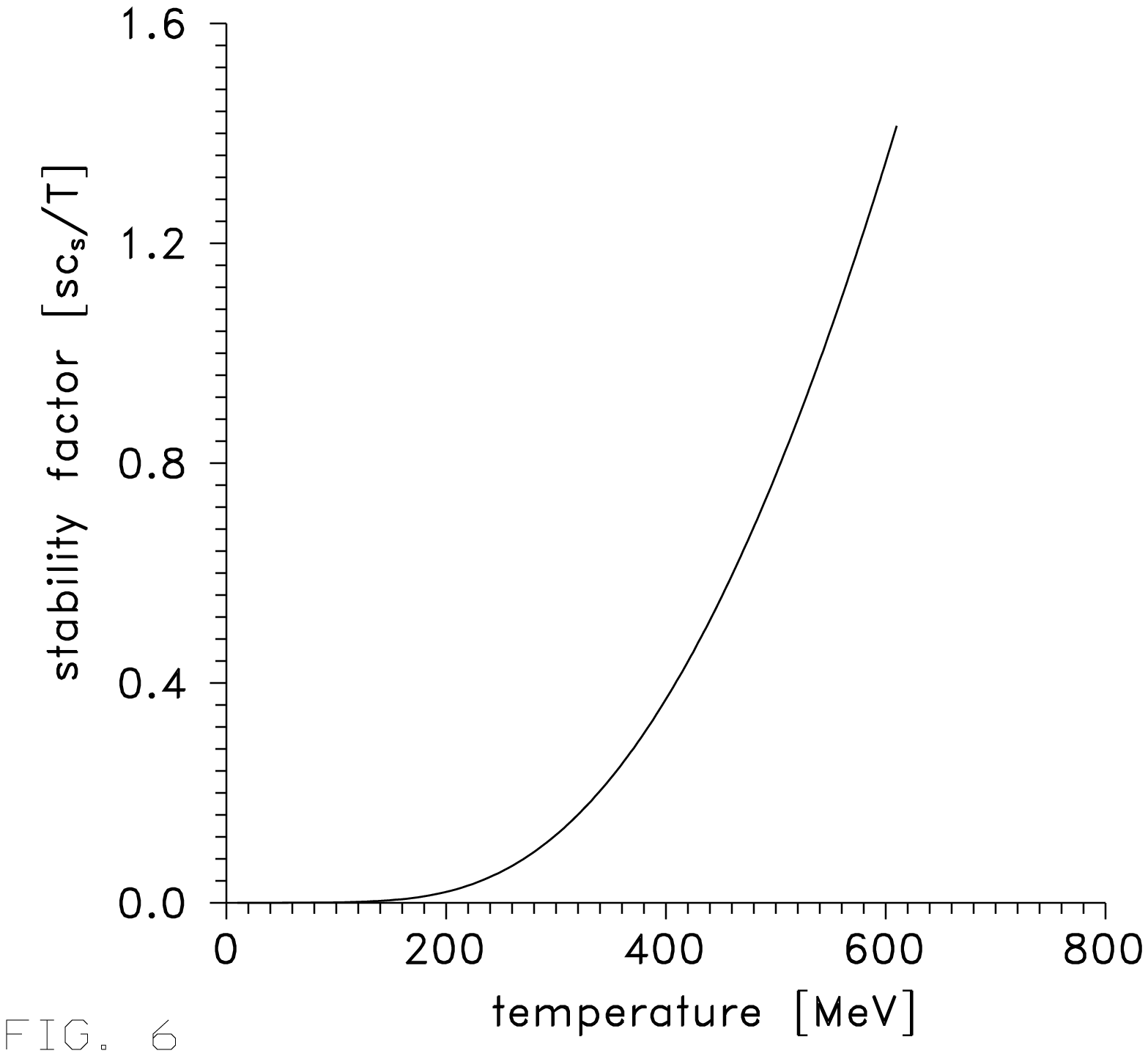}}

\end{document}